\documentclass[cits]{PoS}

\pdfoutput=1

\usepackage{verbatim,dsfont}

\def\tr{\mbox{tr }}
\def\P{\mathcal{P}}

\def\ts{\tilde{\Sigma}}

\title{Dual condensate, dressed Polyakov loops and center symmetry from Dirac spectra}

\ShortTitle{Dual condensate, dressed Polyakov loops ...}

\author{\speaker{Falk Bruckmann} and Christian Hagen\\
        Institute for Theoretical Physics, University of Regensburg, 
        D-93040 Regensburg, Germany}

\author{Erek Bilgici and Christof Gattringer\\
        Institut f\"ur Physik, FB Theoretische Physik, Universit\"at Graz,
        A-8010 Graz, Austria\\ \\
        E-mails:
        \email{erek.bilgici@uni-graz.at, falk.bruckmann@physik.uni-r.de,
        christof.gattringer@uni-graz.at, christian.hagen@physik.uni-r.de}
        }

\abstract{
We construct a novel observable for finite temperature QCD that relates confinement and chiral symmetry.
It uses phases as boundary conditions for the fermions. 
We discuss numerical and analytical aspects of this observable,
like its spectral behavior below and above the critical temperature,
as well as the connection to chiral condensate,
center symmetry and the canonical ensemble.
}

\FullConference{The XXVI International Symposium on Lattice Field Theory \\
                 July 14 - 19, 2008\\
                 Williamsburg, Virginia, USA}

\begin{document}

\section{Motivation and introduction}

The main phenomena in QCD at the finite temperature transition
are deconfinement and chiral symmetry restoration.
There has always been the question whether there is a mechanism connec\-ting the two,
in particular since in the quenched theory the corresponding phase transitions
occur at the same critical temperatures \cite{kogut:83}.
Here we define an observable that indeed links the two
and discuss numerical findings from quenched lattice configurations
\cite{bilgici:08}.

The Polyakov loop (with $\P$ denoting path ordering)
\begin{equation}
 P(\vec{x})\equiv\P \exp\Big(i\int_0^\beta \!\!d x_0 \,\,A_0(x_0,\vec{x})\Big)\,,
\quad \beta=1/k_BT\,, 
\end{equation}
is the order parameter of confinement, being traceless in the confined phase and
moving towards the center of the gauge group, for SU(3) the three elements
$\{1,\exp(2\pi i/3),\exp(4\pi i /3)\}\,\mathds{1}_3$, at temperatures above $T_c$.
This behavior can be understood by its relation to ($\exp(-\beta F)$ of)
the free energy $F$ of a single heavy quark, which is infinite in the confining regime. 

The spectral density of the Dirac operator at the origin, $\rho(0)$, 
on the other hand,
is the order parameter of chiral symmetry. 
It is related to the condensate by the famous Banks-Casher relation~\cite{banks:80}
$\langle\bar{\psi}\psi\rangle=-\pi\rho(0)$ and vanishes above $T_c$.

How does confinement leave a trace in the Dirac spectrum? 
After all, the quarks should not only know about chiral symmetry,
but also that they are (de)confined. 
The answer will lie in the dependence on temporal boundary conditions, as we will show now.

\section{Idea, derivation and interpretation of the new observable}

We use the lattice as a regulator.
The (untraced) Polyakov loop
\begin{equation}
 P(\vec{x})\equiv\prod_{\tau=1}^{N_0}U_0(\tau,\vec{x})\,,
\end{equation}
is built from temporal links.
For the lattice Dirac operator we use the staggered one \cite{kogut:74a}
\begin{equation}
 D(x,y)\equiv\frac{1}{2a}\,\sum_\mu \eta_\mu(x)\big[U_\mu(x)\delta_{x+\hat{\mu},y}-h.c.\big]\,,
\qquad \eta_\mu(x)=(-1)^{x_1+\ldots +x_{\mu-1}}\,,
\label{eq_def_stag}
\end{equation}
which can be viewed as hopping by one link.

It is obvious and well-known that the $k$-th power of the Dirac operator at the same argument,
$D^k(x,x)$, contains all products of links along closed loops of length $k$, 
starting and ending at $x$.
The Polyakov loop is among these loops for $k=N_0$, 
but how to distinguish it from `trivially closed' loops 
(like, e.g., the plaquette), that do not wind around the temporal direction?

The tool for this has been introduced by one of us in \cite{gattringer:06b}. 
One needs phase boundary conditions for the fermions
\begin{equation}
 \psi(x_0+\beta,\vec{x})=e^{i\varphi} \psi(x_0,\vec{x})\,,\qquad \varphi\in[0,2\pi]\,.
\end{equation}
The physical case of antiperiodic fermions is obtained for $\varphi=\pi$.
These boundary conditions\footnote{ 
Note that fermion bilinears like $\psi^\dagger\psi$ are strictly periodic.} 
amount to an imaginary chemical potential. 
They can be easily implemented by replacing the temporal links $U_0$
in some time slice by $e^{i\varphi}U_0$ and likewise
$U_0^\dagger$ by $e^{-i\varphi}U_0^\dagger$.
As a consequence, all Polyakov loops get a factor $e^{i\varphi}$, 
inverse Polyakov loops a factor $e^{-i\varphi}$ and those with higher winding number 
get the corresponding power of that phase factor,
while the trivial loops stay the same.

In this way, the Polyakov loop can be reconstructed from the Dirac spectrum 
by using at least three boundary conditions, see \cite{bruckmann:06b}.
This `thin' Polyakov loop, however, has poor renormalization and scaling properties
and it turned out that in this approach it is UV dominated \cite{bruckmann:06b}.

Influenced by the Jena group \cite{synatschke:07a}, we instead consider the propagator 
with some probe mass $m$.
The Dirac operator at a particular boundary condition $\varphi$ is denoted by $D_\varphi$ 
and we use a geometric series to represent the propagator,
\begin{equation}
 \tr\,\frac{1}{m+D_{\varphi}}=
  \frac{1}{m}\sum_{k=0}^\infty\frac{(-1)^k}{m^k}\,\tr \big[(D_\varphi)^k\big]\,.
\end{equation}
This representation obviously contains all powers of the Dirac operator.
Plugging in the definition (\ref{eq_def_stag}) and the factors of $e^{i\varphi}$,
the propagator is given as a product of links along all closed loops,
\begin{equation}
 \tr\,\frac{1}{m+D_{\varphi}}=
  \frac{1}{m}\sum_{{\rm loops}\:\: l}
  \frac{{\rm sign}(l)}{(2am)^{|l|}}\:
e^{\,i\varphi q(l)}\:\,
\tr_{\!c}\!\!\!\!\prod_{(x,\mu)\in l} U_\mu(x)\,,
\label{eq_pre_formula}
\end{equation}
where $|l|$ is the length of the loop and ${\rm sign}(l)$ comes from the staggered factor.
The ordered product of $U_\mu(x)$ is over all links $(x,\mu)$ in the loop.

Of importance in (\ref{eq_pre_formula}) is the phase factor, where 
$q(l)$ counts how many times the loop winds around the temporal direction.
One can project onto a particular winding $q$ by a Fourier transform w.r.t. $\varphi$,
\begin{equation}
 \frac{1}{2\pi}\int_0^{2\pi} d\varphi \, e^{-i\varphi q}\,.
\end{equation}
Specifying to a single winding, $q=1$, like for the Polyakov loop, we arrive at \cite{bilgici:08}
\begin{equation}
 \ts\equiv
 \int_0^{2\pi} \frac{d\varphi}{2\pi} \, e^{-i\varphi}
 \frac{1}{V}\Big\langle \tr \frac{1}{m+D_\varphi}
 \Big\rangle = 
 \frac{1}{mV}
\sum_{q(l)=1}
 \frac{{\rm sign}(l)}{(2am)^{|l|}}\:
\Big\langle\tr_{\!c}\!\!\!\!\prod_{(x,\mu)\in l} U_\mu(x)\Big\rangle\,.
\label{eq_the_formula}
\end{equation}
This completes the derivation of our new observable $\ts$ ,
which we refer to as the `dual condensate',
because it is obtained through a Fourier transform
from the trace of the propagator.
Indeed, in the massless limit (after the infinite volume limit as usual) 
we obtain the chiral condensate 
\begin{equation}
 \lim_{m\to 0}\lim_{V\to\infty}\ts
=-\int_0^{2\pi} \frac{d\varphi}{2\pi} \,
e^{-i\varphi}
\lim_{m\to 0}\lim_{V\to\infty}\langle\bar{\psi}\psi\rangle_\varphi
=\int_0^{2\pi} \frac{d\varphi}{2} \,
e^{-i\varphi}
\rho(0)_\varphi\,,
\end{equation}
integrated with a phase factor over the boundary conditions.
Making use of the Banks-Casher relation at every individual angle $\varphi$,
we furthermore obtain the representation in terms of the eigenvalue density 
$\rho(0)_\varphi$. 

The right hand side of (\ref{eq_the_formula}) represents the `dressed Polyakov loop', 
that is the set of all loops
which wind once around the temporal direction. 
In the infinite mass limit, detours become suppressed and only the thin, straight 
Polyakov loop survives as it is the shortest possible loop in this set.

We would like to stress that Eq.\ (\ref{eq_the_formula}) is an exact relation 
and is valid for individual configurations.

\begin{figure}[t]
 \begin{center}
 \includegraphics[width=0.48\linewidth]{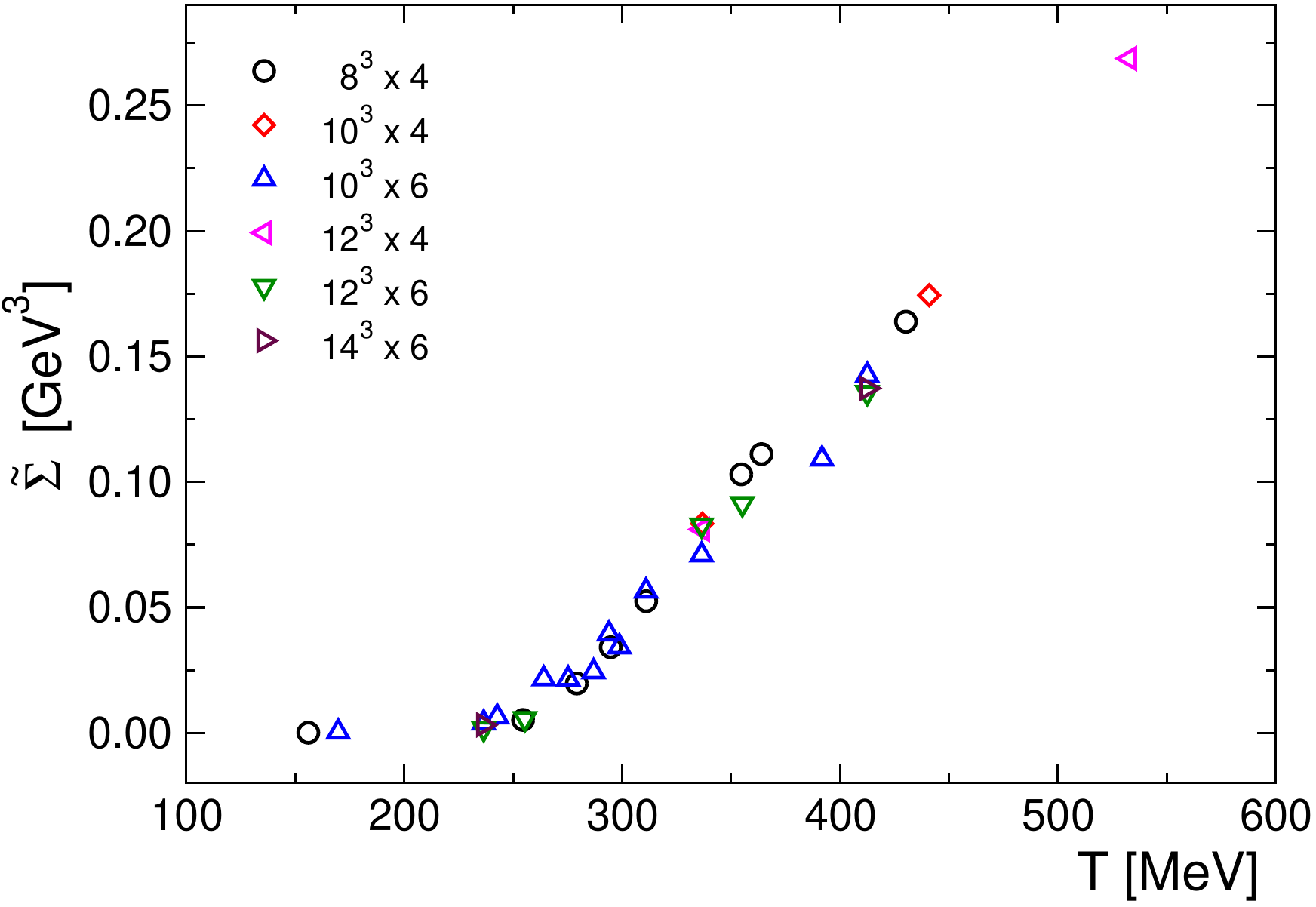}
 \includegraphics[width=0.48\linewidth]{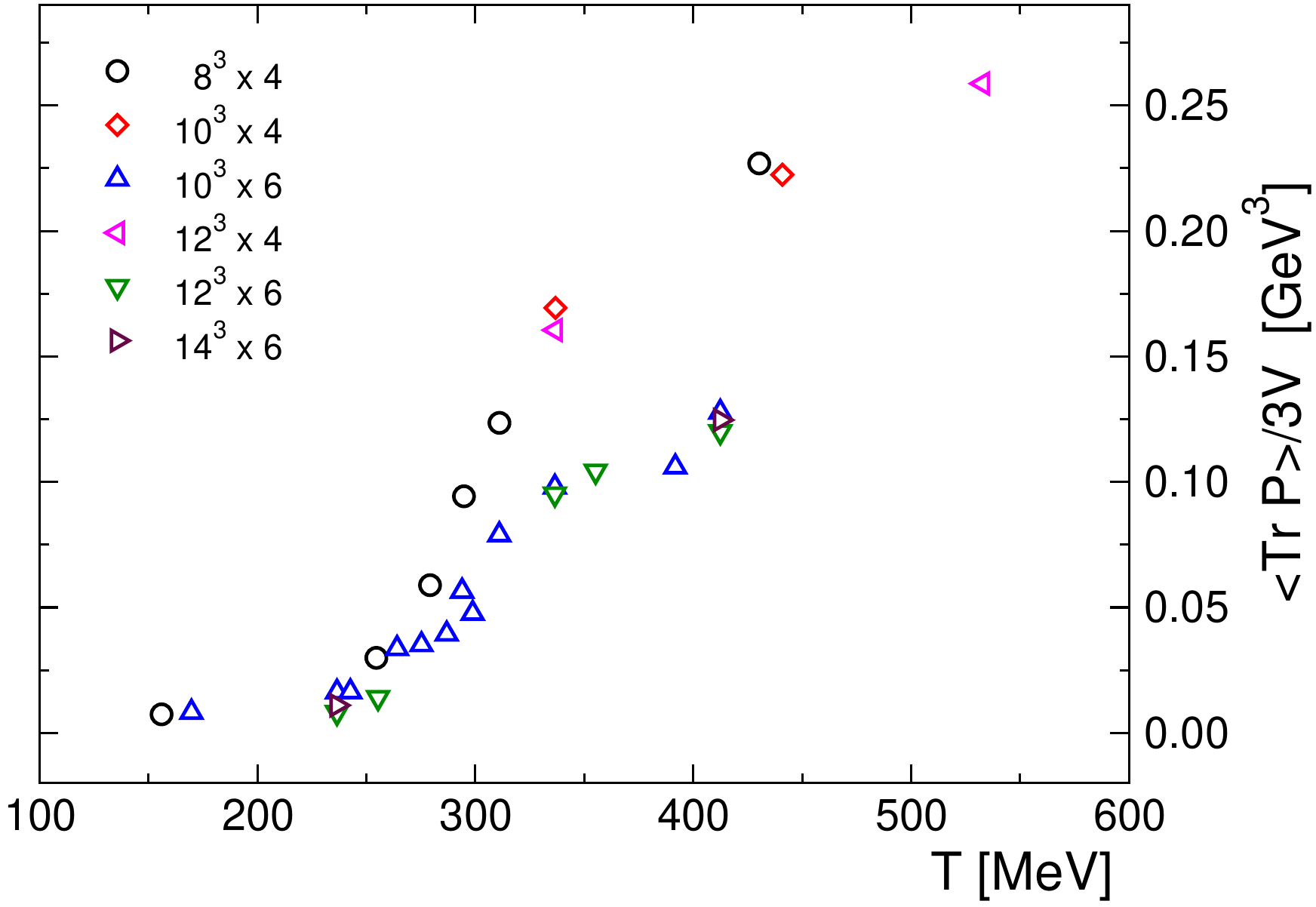}
 \end{center}
 \caption{The dressed Polyakov as an order parameter.
Left: Expectation values for $m =$ 100 MeV as a function of $T$ 
at various lattices, i.e., for different volumes and lattice spacings.
Right: The corresponding plot for the conventional Polyakov loop.}
\label{fig_order_parameter}
\end{figure}

\section{Numerical results and more interpretation}

In the following we discuss various aspects of the relation (\ref{eq_the_formula}).
First of all, Fig.~\ref{fig_order_parameter} shows
that $\ts$ is indeed an order parameter. 
Keeping the mass $m$ fixed, $\ts$ vanishes 
below the critical temperature 
(which is about 280 MeV in the quenched case)
and develops an expectation value for higher temperatures.
One finds that the results (when expressed in physical units)
obtained for different volumes 
and with different resolution 
essentially fall on a universal curve.
This illustrates the good renormalization properties of our observable,
which are inherited from the renormalization of the chiral condensate. 
The improved renormalization properties may also be understood as an effect of
the dressing which renders the new observable less UV dominated.
For comparison we have plotted the corresponding expectation values
of the conventional thin Polyakov loop 
in the right hand side panel of Fig.~\ref{fig_order_parameter}.

\begin{figure}[t]
 \begin{center}
 \includegraphics[width=0.7\linewidth]{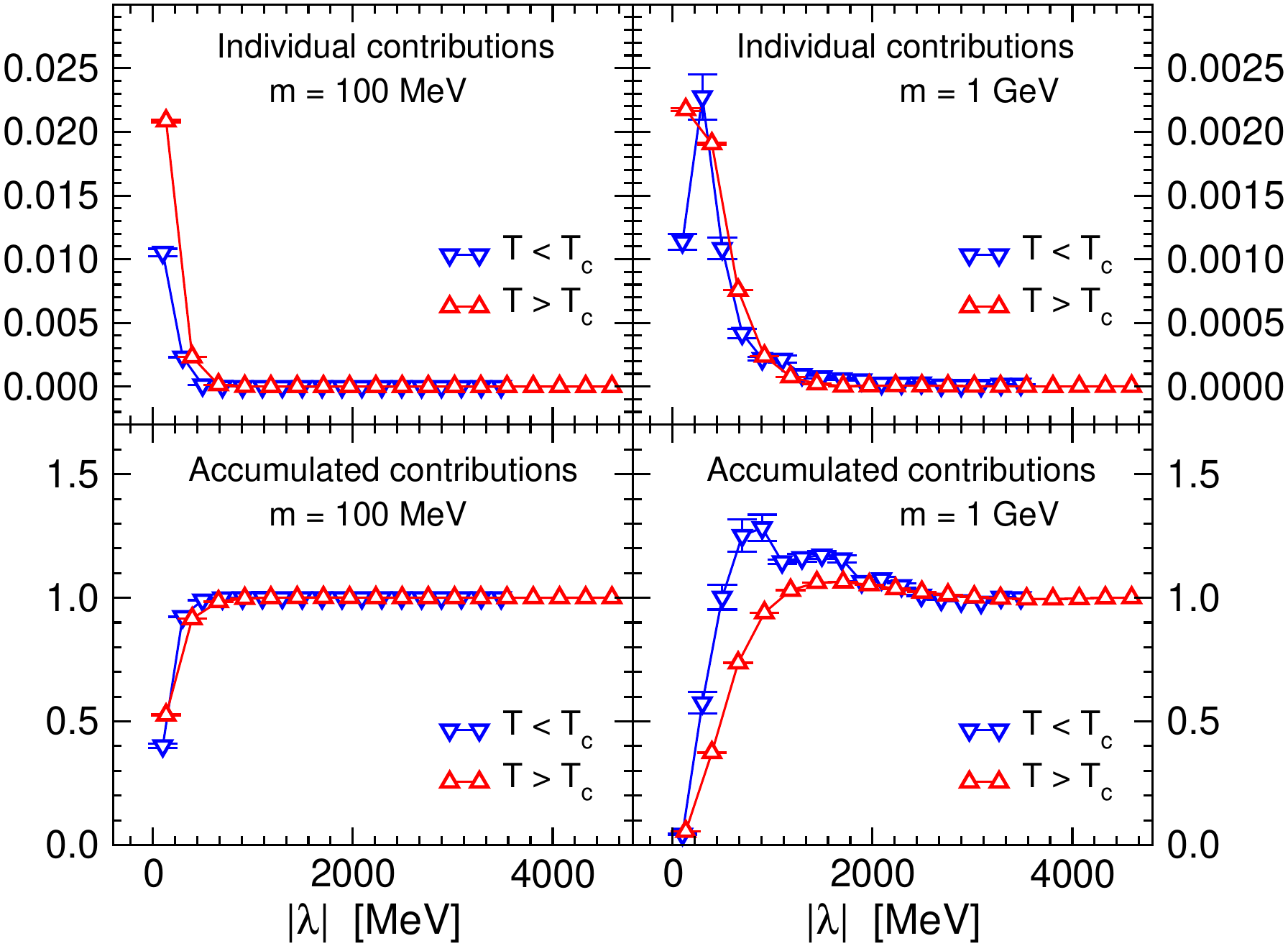}
 \end{center}
 \caption{Individual and accumulated contributions 
 to the spectral sum (\protect\ref{eq_spectral_rep}) at two different masses.}
 \label{fig_spectral_contributions}
\end{figure}

In a spectral representation, the Dirac operator $D_\varphi$ in 
$\ts$ can simply be replaced by a sum over all its eigenvalues, 
\begin{equation}
\ts=\int_0^{2\pi} \frac{d\phi}{2\pi} \, e^{-i\phi}
 \frac{1}{V}\Big\langle \sum_i 
\frac{1}{m+\lambda_\phi^{(i)}}\Big\rangle\,,
\label{eq_spectral_rep}
\end{equation}
where the index $\varphi$ on the eigenvalues $\lambda_\varphi^{(i)}$ 
again refers to the boundary condition angle.
This is actually a suitable representation to numerically compute the dressed Polyakov loop
(in contrast, the right hand side of (\ref{eq_the_formula}) contains an infinite sum over loops
even on a  finite lattice).
On our lattices we have calculated all eigenvalues 
and approximated the $\varphi$-integral by the trapezoidal rule 
with 16 equidistant boundary conditions.

As the eigenvalues appear in the denominator, 
we expect the sum to be dominated by the IR modes.
As Fig.~\ref{fig_spectral_contributions} shows, this is confirmed by the lattice data, 
if $m$ is not too large.

\begin{figure}[b]
 \begin{center}
 \includegraphics[width=0.5\linewidth]{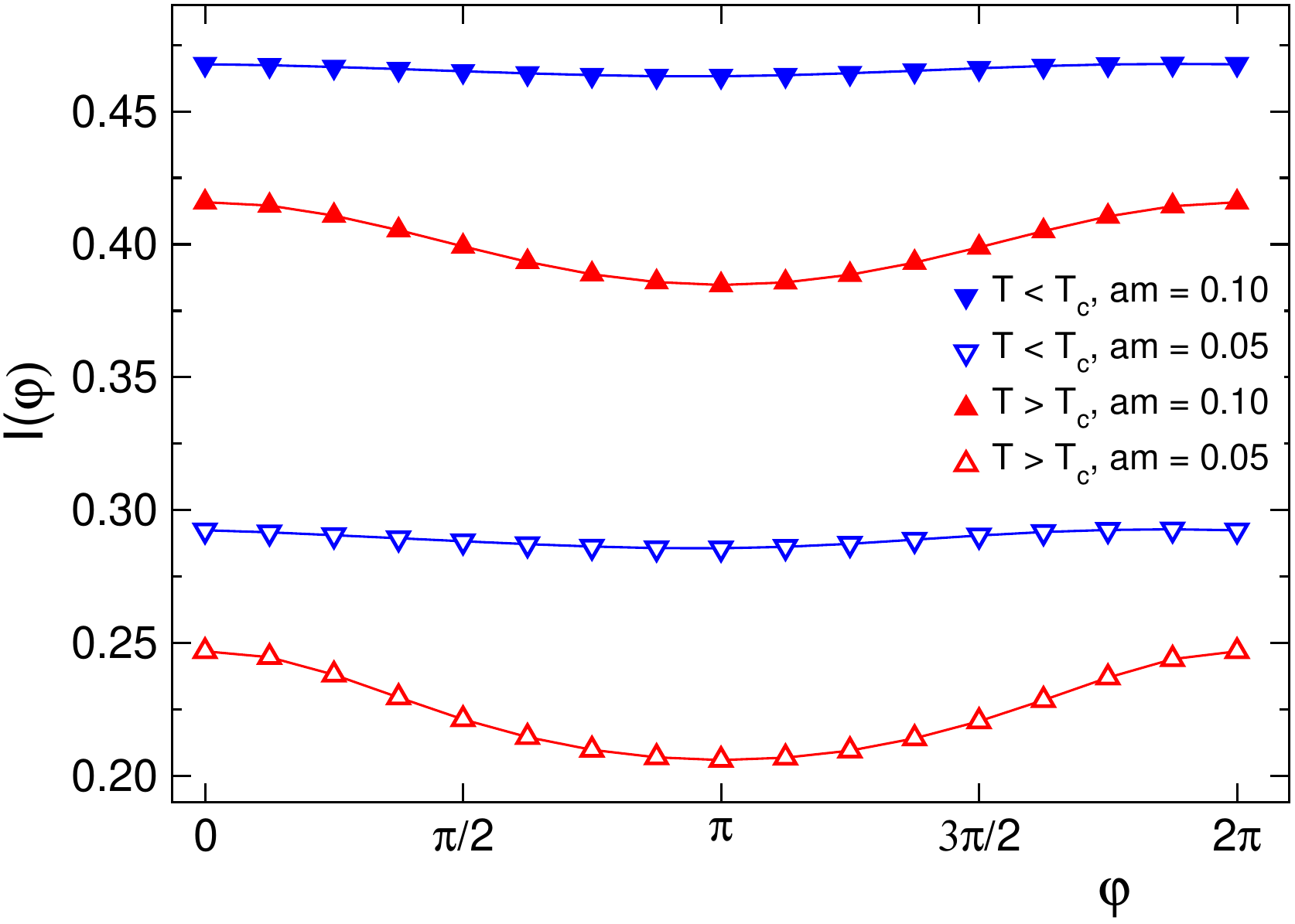}
 \end{center}
 \caption{The integrand $\langle \sum_i (m+\lambda_\varphi^{(i)})^{-1}\rangle/V$
as a function of the boundary condition $\varphi$ for two values of $am$ 
and two temperatures,
 for configurations with real Polyakov loop.
For non-real Polyakov loop the plot is shifted by $\pm 2\pi/3$ restoring the 
$2\pi/3$-symmetry well-known from imaginary chemical potential.}
\label{fig_response_integrand}
\end{figure}

How is a finite resp.\ vanishing order parameter $\ts$ 
built up by the eigenvalues? 
Fig.~\ref{fig_response_integrand} shows that they respond differently to the
boundary conditions in the confined vs.\ deconfined phase.
In that figure we plot the expectation value of the propagator,
i.e., the integrand of $\ts$ without the Fourier factor.
The eigenvalues are independent of the boundary condition in the confined phase, 
which leads to a vanishing order parameter $\ts$. 
In the deconfined phase, on the other hand, 
the eigenvalues show a typical cosine-type of modulation.
Together with the Fourier factor this yields a nonvanishing $\ts$ 
(proportional to the amplitude of the modulation).

The chiral condensate has to behave essentially in the same way, 
as it is the integrand in the massless limit.
Although the chiral condensate is finite in the confined phase, 
it is independent of the boundary condition $\varphi$ 
and hence results in a vanishing dual condensate. 
This feature reflects the conserved center symmetry: 
the trace of the Polyakov loop is zero 
and does not prefer any direction in the space of boundary conditions.

The situation in the deconfined phase might be more confusing at first glance
as the spectrum has a gap there:
how can the vanishing chiral condensate then generate a finite order parameter $\ts$?
The answer is again the dependence on the boundary conditions.
For boundary conditions in line with the original Polyakov loop,
the chiral condensate persists above the critical temperature $T_c$. 
This has been demonstrated on the lattice first in \cite{gattringer:02b} 
and was recently confirmed for gauge group $SU(2)$ in \cite{bornyakov:08a} 
(see also \cite{stephanov:96} for a random matrix study).
This mechanism ensures a finite $\ts$ and should actually be at work for all $T>T_c$.

In the quenched case the conventional Polyakov loop is the order parameter for center symmetry.
Under center transformations the dressed Polyakov loop behaves in the same way.
Therefore, $\ts$ is an order parameter for center symmetry, which is underneath our numerical findings.
In Ref.~\cite{synatschke:08a} it has been proved that all functions of $D_\varphi$ 
integrated with the Fourier factor are center symmetry order parameters and, what is more,
have a well-defined continuum limit.

With dynamical quarks, the center symmetry is broken by the fermion determinant.
One might then expect a $\varphi$-dependence and a nonvanishing $\ts$ also in the low-temperature phase
(and neither the thin nor the dressed Polyakov loop are order parameters in the strict sense).
 
When evaluated in the canonical ensemble with quark number $Q=1$ (see,
e.g., \cite{kratochvila:04}\footnote{
We thank Philippe de Forcrand and Michael Ilgenfritz for discussions on this point.}), 
our observable has an interesting
interpretation. It is obtained as the derivative of the free energy in
the $Q=1$ sector with respect to the mass parameter $m$. In the confining
phase this free energy is infinite and thus independent of $m$, as
suggested by our quenched analysis. Above the transition the free energy
becomes finite and $m$ dependent, consistent with the quenched data.

\section{Summary and outlook}

We have shown that the response of Dirac spectra to different temporal boundary conditions 
contains information about confinement. 
We have defined the dressed Polyakov loop $\ts$ as a novel deconfinement order parameter,
that interpolates between the dual chiral condensate and the conventional thin Polyakov loop
in the different extremes of the mass parameter $m$.
Among other properties we have shown that this quantity is IR dominated 
and transforms non-trivially under center symmetry transformations
(therefore, we expect the findings to be independent of the choice of the lattice Dirac operator).
In the same spirit, 
many center sensitive functions of the Dirac operator can be defined, also in the continuum.

In full QCD (or with the recently studied 4-fermion interactions \cite{sinclair:08}),
the critical temperatures of deconfinement and chiral restoration could be different.
It would be interesting to see what happens then in our formalism. 
As a speculative scenario, a deconfining and chirally broken phase 
would have both the dual and the antiperiodic chiral condensate finite.
This is easily possible, e.g.\ if the condensate has a sine-behavior with the boundary condition.
Even a phase that is confining and chirally symmetric could be realized, 
for instance by a vanishing condensate for all boundary conditions.

\pagebreak
 

\begin{thebibliography}{99}

\bibitem{kogut:83}
J.~Kogut, M.~Stone, H.~W. Wyld, W.~R. Gibbs, J.~Shigemitsu, S.~H. Shenker and
  D.~K. Sinclair, {\it Deconfinement and chiral symmetry restoration at finite
  temperature in SU(2) and SU(3) gauge theories},  {\em Phys.~Rev.~Lett.} {\bf
  50} (1983) 393.

\bibitem{bilgici:08}
E.~Bilgici, F.~Bruckmann, C.~Gattringer and C.~Hagen, {\it {Dual quark
  condensate and dressed Polyakov loops}},  {\em Phys. Rev.} {\bf D77} (2008)
  094007 [arxiv: \href{http://arXiv.org/abs/0801.4051}{{\tt 0801.4051}} [hep-lat]].

\bibitem{banks:80}
T.~Banks and A.~Casher, {\it Chiral symmetry breaking in confining theories},
  {\em Nucl.~Phys.} {\bf B169} (1980) 103.

\bibitem{kogut:74a}
J.~B. Kogut and L.~Susskind, {\it {Hamiltonian Formulation of Wilson's Lattice
  Gauge Theories}},  {\em Phys.~Rev.} {\bf D11} (1975) 395.

\bibitem{gattringer:06b}
C.~Gattringer, {\it Linking confinement to spectral properties of the {D}irac
  operator},  {\em Phys. Rev. Lett.} {\bf 97} (2006) 032003
  [\href{http://arXiv.org/abs/hep-lat/0605018}{{\tt hep-lat/0605018}}].

\bibitem{bruckmann:06b}
F.~Bruckmann, C.~Gattringer and C.~Hagen, {\it Complete spectra of the {D}irac
  operator and their relation to confinement},  {\em Phys.~Lett.} {\bf B647}
  (2007) 56--61 [\href{http://arXiv.org/abs/hep-lat/0612020}{{\tt
  hep-lat/0612020}}].

\bibitem{synatschke:07a}
F.~Synatschke, A.~Wipf and C.~Wozar, {\it Spectral sums of the {D}irac-{W}ilson
  operator and their relation to the {P}olyakov loop},  {\em Phys. Rev.} {\bf
  D75} (2007) 114003 [\href{http://arXiv.org/abs/hep-lat/0703018}{{\tt
  hep-lat/0703018}}].

\bibitem{gattringer:02b}
C.~Gattringer and S.~Schaefer, {\it New findings for topological excitations in
  {$SU(3)$} lattice gauge theory},  {\em Nucl.~Phys.} {\bf B654} (2003) 30--60
  [\href{http://arXiv.org/abs/hep-lat/0212029}{{\tt hep-lat/0212029}}].

\bibitem{bornyakov:08a}
V.~G. Bornyakov {\em et.~al.}, {\it {The topological structure of SU(2)
  gluodynamics at T > 0 : an analysis using the Symanzik action and Neuberger
  overlap fermions}},  arxiv: \href{http://arXiv.org/abs/0807.1980}{{\tt 0807.1980}} [hep-lat],
T.~G. Kovacs, 
{\em PoS} {\bf LAT2008} (2008) 198.

\bibitem{stephanov:96}
M.~A. Stephanov, {\it {Chiral symmetry at finite T, the phase of the Polyakov
  loop and the spectrum of the Dirac operator}},  {\em Phys. Lett.} {\bf B375}
  (1996) 249--254 [\href{http://arXiv.org/abs/hep-lat/9601001}{{\tt
  hep-lat/9601001}}].

\bibitem{synatschke:08a}
F.~Synatschke, A.~Wipf and K.~Langfeld, {\it {Relation between chiral symmetry
  breaking and confinement in YM-theories}},  {\em Phys. Rev.} {\bf D77} (2008)
  114018 [arxiv: \href{http://arXiv.org/abs/0803.0271}{{\tt 0803.0271}} [hep-lat]].

\bibitem{kratochvila:04}
S.~Kratochvila and P.~de~Forcrand, {\it {QCD at small baryon number}},  {\em
  Nucl. Phys. Proc. Suppl.} {\bf 140} (2005) 514--516
  [\href{http://arXiv.org/abs/hep-lat/0409072}{{\tt hep-lat/0409072}}],
A.~Alexandru, M.~Faber, I.~Horvath and K.-F. Liu, {\it Lattice QCD at finite
  density via a new canonical approach},  {\em Phys. Rev.} {\bf D72} (2005)
  114513 [\href{http://arXiv.org/abs/hep-lat/0507020}{{\tt hep-lat/0507020}}],
S.~Kratochvila and P.~de~Forcrand, {\it QCD at zero baryon density and the
  Polyakov loop paradox},  {\em Phys. Rev.} {\bf D73} (2006) 114512
  [\href{http://arXiv.org/abs/hep-lat/0602005}{{\tt hep-lat/0602005}}],
P.~de~Forcrand and S.~Kratochvila, {\it Finite density QCD with a canonical
  approach},  {\em Nucl. Phys. Proc. Suppl.} {\bf 153} (2006) 62--67
  [\href{http://arXiv.org/abs/hep-lat/0602024}{{\tt hep-lat/0602024}}],
A.~Li, A.~Alexandru and K.-F. Liu, {\it New results using the canonical
  approach to finite density QCD},  {\em PoS} {\bf LAT2007} (2007) 203
  [arxiv: \href{http://arXiv.org/abs/0711.2692}{{\tt 0711.2692}} [hep-lat]],
J.~Danzer and C.~Gattringer, {\it Winding expansion techniques for lattice QCD
  with chemical potential},  arxiv: \href{http://arXiv.org/abs/0809.2736}{{\tt
  0809.2736}} [hep-lat].

\bibitem{sinclair:08}
D.~K. Sinclair, {\it {Separating the scales of confinement and chiral-symmetry
  breaking in lattice QCD with fundamental quarks}},
  arxiv: \href{http://arXiv.org/abs/0805.4627}{{\tt 0805.4627}} [hep-lat].

\end{thebibliography}

\end{document}